# Angle-Resolved Cryogenic Brillouin-Mandelstam Spectroscopy of Surface and Bulk Acoustic Phonons in Diamond

Jordan Teeter[1,2], Dylan Wright[1,2], Nidhish Thiruthukkal Puthenveettil[1,2], Fariborz Kargar[3,*], and Alexander A. Balandin[1,2,4,*]

[1]Department of Materials Science and Engineering, University of California, Los Angeles, California 90095 USA

[2]California NanoSystems Institute, University of California, Los Angeles, California 90095 USA

[3]Materials Research and Education Center, Department of Mechanical Engineering, Auburn University, Auburn, Alabama 36849 USA

[4]Center for Quantum Science and Engineering, University of California, Los Angeles, California, 90095 USA

We used angle-resolved Brillouin-Mandelstam light-scattering spectroscopy to monitor surface and bulk acoustic phonons in diamond along the <100> and <110> crystallographic directions across a temperature range from 10 K to 300 K. The frequencies and phase velocities were measured for three types of surface acoustic phonons: Rayleigh waves, shear horizontal waves, and high-frequency pseudo-longitudinal waves. All surface acoustic phonons exhibit weak temperature dependence, with the largest observed change of 1.6% across the examined temperature range. The frequencies of all three types of surface acoustic phonons agree with the theoretical values within the experimental uncertainty. Cryogenic surface-acoustic-phonon data are important for diamond-based quantum sensors, surface acoustic wave devices, and other electronic technologies. Knowledge of surface acoustic phonons can also be used for developing accurate models for thermal transport between interfaces.

* Corresponding authors: balandin@seas.ucla.edu and fkargar@auburn.edu

Diamond has emerged as a technologically important material platform for radio-frequency electronics, high-power diodes and transistors, surface acoustic wave (SAW) devices, thermal management, and quantum technologies due to its exceptional mechanical, thermal, and electronic properties[1–9]. In particular, diamond possesses the highest known acoustic phonon velocities among bulk solids, with longitudinal and transverse acoustic velocities of approximately 18 km $s^{-1}$ and 12 km $s^{-1}$, respectively, and a Rayleigh-wave velocity approaching 11 km $s^{-1}$ at room temperature.[15,16] These properties make diamond an attractive material for next-generation high-frequency SAW devices, where the superior acoustic phonon velocities enable operation at increasing frequencies. Surface acoustic phonons are critical for SAW-based signal processing and sensing applications, where device characteristics such as frequency stability, propagation loss, and electromechanical coupling depend directly on surface acoustic phonon velocities and their temperature dependence[10–14].

Beyond "conventional" electronics, diamond is a leading material platform for quantum sensing and quantum information technologies based on nitrogen–vacancy (NV) centers[15]. The figure of merit for NV-based quantum sensors and qubits is the spin phase coherence time, $\tau_2$, which determines the lifetime of superposition for NV centers. One major component that constrains $\tau_2$ is thermal fluctuations from phonons with a characteristic longitudinal relaxation time, $\tau_1$. Uniquely for high-purity diamond, with a Debye temperature of $\theta_D$ = 1880 K[16], the high-frequency acoustic phonons and quasi-localized phonons are dominant drivers of $\tau_1$ from room temperature (RT) to low temperatures[17–19]. Thus, investigating surface and bulk phonons in diamond, particularly at cryogenic temperatures, is essential for quantum technologies[20–22]. Accurate cryogenic phonon data is also required to model thermal boundary resistance (TBR) in diamond-based heterostructures, which is increasingly relevant for quantum devices, cryogenic electronics, and hybrid acoustic–optical platforms. The importance of TBR in cryogenic thermal management increases[23]. Consider Kapitza resistance, $R_K$, which is the fundamental component of TBR, emerging due to the mismatch in acoustic phonons at the interface of two materials. It is commonly defined as $R_K = \Delta T/Q$, where $\Delta T$ is the temperature drop across the interface, and $Q$ is the heat flux through the interface[24]. Kapitza resistance grows as the temperature, $T$, decreases, scaling at low temperatures as $R_K \propto 1/T^3$.[25] The latter explains its significance in the overall thermal resistance of the device structures. Accurate acoustic phonon energies and dispersion are also needed for the phonon engineering approaches[26] and their applications to the thermal optimization of device structures implemented with ultra-wide-band-gap (UWBG) semiconductors[27]. Despite the importance of phonon–spin and phonon–phonon interactions in all types of the above-mentioned devices, experimental data on surface acoustic phonons in diamond at low temperatures remain scarce.

Inelastic light scattering provides a direct, noncontact method to probe acoustic phonons in solids[28]. Angle-resolved Brillouin–Mandelstam spectroscopy (BMS), also referred to as Brillouin light scattering (BLS) spectroscopy, enables selective detection of both bulk and surface acoustic modes via elasto-optic and surface-ripple scattering mechanisms, respectively[28–30]. While bulk acoustic phonons and certain types of acoustic surface phonons in diamond have been studied extensively at room temperature (RT)[31–39], the properties of surface acoustic phonons at cryogenic temperatures still need to be addressed. We are aware of a few reports of the elastic constants of crystalline diamond at cryogenic temperatures[40–42]. No reports of the temperature sensitivity of the Rayleigh wave (RW), shear horizontal wave (SHW), or high-frequency pseudo-longitudinal wave (HFPLW) surface acoustic phonons are available.

In this work, we employ angle-resolved BMS spectroscopy to measure bulk and surface acoustic phonons in single-crystal diamond along the ⟨100⟩ and ⟨110⟩ crystallographic directions over a temperature range from 10 K to 300 K. BMS relies on the inelastic scattering of light by thermally generated phonons, measuring the frequency shift between incident light and scattered light[28]. In the standard backscattering configuration, the technique distinguishes between phonons propagating in the bulk and on the surface of the material, with the probed wavevector of bulk phonons determined by $q_B = 4\pi\eta/\lambda_0$ , where $\eta$ is the refractive index and $\lambda_0$ is the incident wavelength. Phonons propagating on the surface generate surface ripples that scatter light, with the probed wavevector given by $q_S = 4\pi\sin(\theta)/\lambda_0$ where $\theta$ is the incidence angle. While the Rayleigh wave is the primary surface-localized acoustic mode, anisotropic crystals can also support pseudo-surface acoustic waves (PSAWs), or leaky surface waves. In diamond, two high-velocity PSAWs with predominantly longitudinal-like and shear-horizontal-like character are observed along specific crystallographic directions[35,43–45].

To characterize these modes, incidence angle and in-plane rotational Brillouin measurements were performed using a 532 nm laser (Spectra-Physics Excelsior 300 mW). The sample was vertically mounted on a motorized stage, allowing for a full 360° in-plane rotation ($\phi$) and a complete range of incidence angles ($\theta$). Incident light was *s*-polarized and attenuated to 20 mW at the sample surface. Backscattered light was collected without polarization selection and focused through a pinhole into a "3 + 3 pass" tandem Fabry-Perot interferometer (TFP, JRS Scientific Instruments). All spectra were normalized to the central elastic peak. More details about our BMS measurement procedures can be found elsewhere in the context of other materials systems[46,47]. Figures 1 (a, b) show the dependence of the spectral positions of surface acoustic waves, including the shear horizontal wave (SHW) and high-frequency pseudo-longitudinal wave (HFPLW), on the incidence angle $\theta$ along the $\langle 100 \rangle$ and $\langle 110 \rangle$ crystallographic directions, respectively. In both cases, two additional peaks, labeled TA and LA, correspond to transverse and longitudinal acoustic phonons with wavevectors oriented primarily along the surface normal in the [001] direction.

[**Figure 1**: Brillouin spectra of diamond showing surface acoustic waves of shear horizontal wave (SHW) and high-frequency pseudo-longitudinal wave (HFPLW) propagating along (a) $\langle 100 \rangle$, and (b) $\langle 110 \rangle$ crystallographic directions as a function of incidence angle. In both panels, the peaks correspond to bulk transverse acoustic (TA) and longitudinal acoustic (LA) phonons. Note that these bulk phonons have wavevectors oriented approximately along the surface normal, *i.e.*, the [001] direction, while propagating within (100) and (110) planes, respectively. The spectra for different angles are shifted vertically for convenience. All data are for temperature $T$ = 298 K.]

As seen in Figure 1, the SHW and HFPLW frequencies exhibit a strong dependence on $\theta$, whereas the bulk TA and LA modes show minimal variation, as expected. For bulk LA and TA modes, the $q$-vector is primarily determined by the refractive index ($\eta \sim 2.42$ for diamond at 532 nm)[48,49]. The contour plots in Figure 2 (a, b) show the surface waves as a function of $q$-vector for $\langle 100 \rangle$ and $\langle 110 \rangle$ directions, respectively. The dispersive behavior of the SHW and HFPLW modes is similar along both crystallographic directions, although their absolute frequencies differ. For the same wavevectors, the HFPLW exhibits higher frequencies along the $\langle 100 \rangle$ direction and lower frequencies along ⟨110⟩, whereas the SHW shows the opposite trend, with lower frequencies along ⟨100⟩ and higher frequencies along ⟨110⟩ across the full $q$ range. In contrast, the frequency of bulk modes shown in Figures 2 (c, d) remains relatively invariant with $\theta$, consistent with the dependence on refractive index. The bulk TA mode propagating close to the [001] direction exhibits different degeneracy behavior depending on the plane of propagation. Although in both cases, the deviation from [001] is small, the symmetry of the propagation direction differs. When the wavevector lies within the (010) plane (Figure 2 (c)), the two transverse acoustic modes remain degenerate. In contrast, when the wavevector lies within the (110) plane (Figure 2 (d)), the degeneracy is lifted, and the TA modes become nondegenerate. This difference arises from the distinct symmetry properties of the two planes, which determine whether the transverse polarizations remain equivalent.

[**Figure 2**: Brillouin intensity contour maps showing the SHW and HFPLW surface modes along the (a) ⟨100⟩, and (b) ⟨110⟩ directions. Panels (c) and (d) show bulk acoustic modes for wavevectors close to [001] propagating within the (010) and (110) planes. Note that the transverse acoustic modes remain degenerate in the (010) plane and split into two branches in the (110) plane.]

To characterize the elastic anisotropy, Brillouin spectra were collected at a fixed incident angle ($\theta$ = 70°, $q_s$ = 0.022 $nm^{-1}$), while varying the azimuthal angle ($\phi$) in 15° increments, starting from the ⟨100⟩ direction ($\phi$ = 0°). The spectra for the resulting 360° in-plane rotation are presented in Figure 3 (a), where a distinct four-fold symmetry of the SHW and HFPLW peak frequencies is presented in Figs. 3 (b, c). As seen, the frequency extremes of the SHW mode reach a minimum along the ⟨100⟩ axes ($\phi$ = 0°, 90°, 180°, 360°), whereas the HFPLW reaches its maximum. Conversely, along the ⟨110⟩ directions ($\phi$ = 45°, 135°, 225°, 315°), the SHW is maximum, while the HFPLW is minimal. For the SHW, the frequencies fluctuate between 46.5 GHz ($v_p$ = 13.16 km/s) and 42.2 GHz ($v_p$ = 11.9 km/s). The HFPLW exhibits a similar range from 66.38 GHz ($v_p$ = 18.7 km/s) to 63.56 GHz ($v_p$ = 17.9 km/s). These results show excellent agreement with established room-temperature data[35].

[**Figure 3**: Brillouin spectra of diamond for (a) 360° in-plane rotation, probing phonons from ⟨100⟩ in increments of 15° at a fixed incidence angle of 70°. Panels (b, c) show polar plots of the SHW and HFPLW anti-Stokes peak frequencies as a function of in-plane rotation angle, $\phi$. All data are for temperature $T$ = 298 K.]

After establishing baseline velocities and comparing to RT literature values, low temperature measurements were performed on a separate BMS system, equipped with a helium cryostat (Montana Instruments). The sample was mounted vertically using (VGE-7031 Varnish) to ensure minimal sample drift and good thermal contact across temperatures. Optical excitation was provided by a 532 nm laser (Coherent Verdi V-2) focused through a long working distance objective (Mitutoyo 10×, NA = 0.28). The beam path traversed two windows (N-BK7, vacuum shroud, and radiation shield), resulting in an effective working distance of approximately 35 mm and an RMS spot radius of 850 nm. Incident power was maintained at 6.5 mW to ensure negligible laser-induced heating. The backscattered light was focused through a pinhole into a JRS Scientific Instruments triple pass tandem Fabry-Perot interferometer (TFP-HC2). The polarization of the incident light in this work is $s$-polarized, with the interferometer selecting for $s$-polarized light, resulting in straight-polarized light being analyzed for these experiments. All experiments were performed on the same diamond sample for the ⟨100⟩ and ⟨110⟩ directions.

Figures 4 (a, b) present the BMS spectra of surface acoustic phonons in diamond along the ⟨100⟩ and ⟨110⟩ crystallographic directions, respectively. In both cases, the bulk TA and LA phonons have wavevectors with $q_B$ = 0.057 nm$^{-1}$ with directions close to [001], but propagate within different planes, namely (010) and (110), respectively. The temperature of the measurement spans from ~300 K down to ~10 K. Five distinct acoustic phonon modes are resolved and identified. They are labeled as RW, SHW, HFPLW, TA, and LA. All measurements were conducted at a fixed incidence angle ($\theta$ = 55°, $q_S$ = 0.019 nm$^{-1}$) and normalized to the elastic centerline for comparison. Figure 4 (c, d) show the normalized anti-Stokes peak intensities of the SHW and LA modes for $\phi$ = 0° and $\phi$ = 45°, respectively, measured from 300 K down to ~10 K. As indicated by the fitted lines, the BMS signal intensity decreases approximately linearly with temperature for both modes.

In semiconductors, the BMS intensity is primarily defined by the Bose–Einstein population of acoustic phonons, *i.e.*, $I \propto n(\omega, T) = (\exp[\hbar\omega_p/k_bT] - 1)^{-1}$, where $n$ is the phonon population, $k_b$ is the Boltzmann constant, $\hbar$ is the reduced Planck constant, and $\omega_p$ is the angular phonon frequency. In the high-temperature limit, $k_bT \gg \hbar\omega_p$, this expression reduces to $n \cong k_bT/\hbar\omega_p$, leading to an approximately linear temperature dependence of the BMS intensity. Even at the lowest measured temperature, $T = 9.3$ K, $k_bT$ remains comparable to the energy of the highest-frequency phonon mode, *i.e.*, the [001] LA peak at $f \sim 160$ GHz. Therefore, no strong nonlinearity in the BMS intensity is expected within the measured temperature range.

[**Figure 4**: BMS spectra of diamond probing phonons propagating in the (a) ⟨100⟩, and (b) ⟨110⟩ crystallographic directions showing Rayleigh wave (RW), shear horizontal wave (SHW), and high-frequency pseudo-longitudinal wave (HFPLW). In both panels, the peaks corresponding to bulk transverse acoustic (TA) and longitudinal acoustic (LA) phonons are observed, with phonon wavevectors oriented approximately along the surface normal, *i.e.*, the [001] direction. The spectra for different temperatures are shifted vertically for convenience. The peak at ~28 GHz indicated with (*) is a spectral artifact from the cryostat window. (c, d) Temperature-dependent anti-Stokes peak intensities of the SHW and LA phonons for (c) ⟨100⟩, and (d) ⟨110⟩ orientations. Linear fits to the SHW and LA modes are shown, along with the corresponding $R^2$ values, indicating an approximately linear temperature dependence.]

Figure 5 (a) presents the experimental BMS frequency shift and the calculated phase velocity for the surface acoustic phonons along the ⟨100⟩ and ⟨110⟩. For the ⟨100⟩ direction (Figure 5 (a)), phase velocities for HFPLW, SHW, and RW vary by only 0.4%, 0.3%, and 0.9%, respectively. Correspondingly, the ⟨110⟩ direction shows limited thermal variation with modes in this orientation displaying shifts of 1.0% (HFPLW), 0.4% (SHW), and 1.6% (RW) over the cryogenic range. Figure 5 (b) shows bulk modes in the [001] propagating in the (010) and (110) planes. The corresponding bulk LA and TA modes in the (010) plane show negligible temperature dependence upon cooling from 300 K to 10 K. The LA mode exhibits a frequency shift of only 0.2%, whereas the bulk TA mode changes by just 0.01% between 300 K and 100 K before becoming indistinguishable from the background. Similarly, the bulk LA mode and TA mode propagating in the (110) plane close to the [001] direction shifted only by ~0.7% and 0.2%, respectively. This weak temperature dependence of both bulk LA, TA phonons and surface HFPLW, SHW, RW modes reflects the stability of the underlying elastic constants governing these excitations. These observations agree with the temperature invariance of $C_{11}$, $C_{12}$, and $C_{44}$ parameters, as has been previously observed in monocrystalline diamond, aligning with expectations[40–42]. The calculated phase velocities for all observed phonon modes are summarized in Table I for ⟨100⟩ and Table II for ⟨110⟩.

[**Figure 5**: (a) Measured surface acoustic phonon frequency and velocity for diamond in the ⟨100⟩ and ⟨110⟩ directions. (b) Bulk acoustic phonon frequency and velocity for the corresponding directions [001] propagating in the (010) and (110) planes.]

As mentioned in the introduction, heat transport at the solid-solid interface is a critical issue in modern engineering applications. The conventional approach for evaluating TBR is based on two prevailing theories – the acoustic mismatch model (AMM) and the diffuse mismatch model (DMM)[25]. In the AMM, phonons are treated as elastic waves incident on an ideal, atomically smooth interface, and their transmission and reflection are determined by the acoustic impedance mismatch between the two materials. In contrast, the DMM assumes that phonons are scattered diffusely at the interface so that their transmission probability is governed by the relative phonon densities of states and velocities in the two media. Both approaches emphasize the central role of acoustic phonons in determining the thermal boundary conductance, $G = 1/R_K$, which can be expressed as an integral over phonon modes, $G \sim \int C(\omega)\, v_g(\omega)\, \alpha(\omega)\, d\omega$, where $v_g$ is the phonon group velocity and $\alpha(\omega)$ is the transmission coefficient across the interface. Within AMM, which is more relevant for cryogenic temperatures, $\alpha(\omega)$ is governed by the acoustic impedance $Z = \rho v_g$, leading to a strong dependence of $G$ on phonon velocities, often approximated as $G \propto 1/v_g^2$ for mismatched interfaces. Consequently, even modest variations in acoustic phonon velocities and dispersion can influence interfacial transmission and heat flow, particularly at cryogenic temperatures where long-wavelength acoustic phonons make a significant contribution to thermal transport. Therefore, accurate knowledge of acoustic phonon velocities at low temperatures is important for the estimation of Kapitza resistance. Further refinement of AAM and DMM likely requires inclusion of the interface or surface phonons.

In conclusion, we used BMS spectroscopy to monitor surface and bulk acoustic phonons in diamond across a temperature range from 10 K to 300 K. The frequencies and phase velocities were measured for three types of surface acoustic phonons: Rayleigh waves, shear horizontal waves, and high-frequency pseudo-longitudinal waves. All surface acoustic phonons are temperature-stable. No deviation from linear scaling with temperature was observed for the normalized intensity of the acoustic phonons. We argued that the knowledge of surface acoustic phonon characteristics is essential for developing accurate models of TBR at cryogenic temperatures. The cryogenic surface acoustic phonon data can be used for optimization of diamond-based quantum sensors, surface acoustic wave devices, UWBD diodes and transistors, and other electronic technologies.

**Acknowledgments**

The work at UCLA was supported, in part, by ULTRA, an Energy Frontier Research Center (EFRC) funded by the U.S. Department of Energy, Office of Science, Basic Energy Sciences under Award # DE-SC0021230. A.A.B. and F.K. acknowledge the support of the National Science Foundation (NSF) *via* a Major Research Instrument (MRI) DMR Project No. 2019056 entitled "Development of a Cryogenic Integrated Micro-Raman-Brillouin-Mandelstam Spectrometer."

**Conflict of Interest**

The authors declare no conflict of interest.

**Author Contributions**

A.A.B. and F.K. conceived the idea and led the data analysis and manuscript preparation. J.T., D.W., and N.T.P performed the Brillouin-Mandelstam spectroscopy and contributed to data analysis. All authors reviewed and contributed to the final manuscript.

**The Data Availability Statement**

The data in support of the findings of this study are available from the corresponding author upon reasonable request.

**Table I: Phonon Phase Velocities for <100> Diamond**

| | <100> | | | [001] → (010) | |
|---|---|---|---|---|---|
| Temperature (K) | RW $v_p$ (km/s) | SHW $v_p$ (km/s) | HFPLW $v_p$ (km/s) | TA $v_p$ (km/s) | LA $v_p$ (km/s) |
| 298 | 10.326 | 12.073<br>12.268[35]<br>12.282[31] | 18.915<br>18.22[35]<br>18.33[31] | 12.548 | 17.827<br>17.44[35]<br>17.51[31] |
| 195 | 10.326 | 12.053 | 19.058 | 12.470 | 17.774 |
| 146 | 10.228 | 12.053 | 19.006 | 12.547 | 17.767 |
| 101 | 9.972 | 12.053 | 19.015 | 12.550 | 17.769 |
| 76 | 10.206 | 12.070 | 18.889 | - | 17.773 |
| 48 | 10.212 | 12.057 | 18.996 | - | 17.758 |
| 27 | 10.202 | 12.096 | - | - | 17.787 |
| 9.3 | 10.225 | 12.112 | - | - | 17.786 |

- Implies peak was indistinguishable from background

**Table II: Phonon Phase Velocities for <110> Diamond**

| | <110> | | | [001] → (110) | |
|---|---|---|---|---|---|
| Temperature (K) | RW $v_p$ (km/s) | SHW $v_p$ (km/s) | HFPLW $v_p$ (km/s) | TA $v_p$ (km/s) | LA $v_p$ (km/s) |
| 298 | 10.417 | 12.884<br>12.268[35]<br>12.282[31] | 17.736<br>18.32[31] | 12.030 | 17.759<br>17.51[31] |
| 240 | - | 12.798 | 18.476 | 12.117 | 17.730 |
| 195 | 10.093 | 12.774 | 17.653 | 12.097 | 17.698 |
| 146 | 10.214 | 12.975 | 18.241 | 12.097 | 17.761 |
| 101 | 10.086 | 12.951 | 18.222 | - | 17.748 |
| 76 | - | 12.989 | 17.544 | 12.056 | 17.759 |
| 48 | 10.042 | 13.086 | - | - | 17.763 |
| 27 | 10.458 | 12.989 | - | - | 17.750 |
| 9.3 | 10.245 | 12.946 | - | - | 17.631 |

- Implies peak was indistinguishable from background

## References

[1] J.T. Glass, B.A. Fox, D.L. Dreifus, and B.R. Stoner, "Diamond for Electronics: Future Prospects of Diamond SAW Devices Electronic-Applications Overview," MRS Bull. **23**(9), 49–55 (1998).

[2] A. Aleksov, M. Kubovic, M. Kasu, P. Schmid, D. Grobe, S. Ertl, M. Schreck, B. Stritzker, and E. Kohn, "Diamond-based electronics for RF applications," Diam. Relat. Mater. **13**(2), 233–240 (2004).

[3] X. Zhang, M. Xu, E.J. Garratt, S. Luo, B.B. Pate, T.S. Pieshkov, A.G. Birdwell, T. Gray, A. Biswas, A.B. Puthirath, M.R. Neupane, T.G. Ivanov, Y. Zhao, R. Vajtai, and P.M. Ajayan, "Diamond Epilayers with Subnanometer Surface Roughness for Enhanced Device Performance," ACS Appl. Electron. Mater. **8**, 961–971 (2026).

[4] S. Ghosh, H. Surdi, F. Kargar, F.A. Koeck, S. Rumyantsev, S. Goodnick, R.J. Nemanich, and A.A. Balandin, "Excess noise in high-current diamond diodes," Appl. Phys. Lett. **120**(6), 062103 (2022).

[5] X. Zhang, C. Chang, Q. Zhu, S. Luo, R. Vajtai, Y. Zhao, and P.M. Ajayan, "Scalable selective-area diamond growth for thermal management applications," Appl. Phys. Lett. **128**(8), 081901 (2026).

[6] Y. Yue, M. Wang, Y. Liu, R. Guo, H. Zhang, H. Xie, Y.S. Ang, and S. Fang, "Effects of hydrostatic compression and tension on silicon-vacancy centers in diamond," Appl. Phys. Lett. **128**(5), (2026).

[7] W. Shen, X. Li, X. Wang, Z. Yin, L. Zhen, H. Wen, Z. Ma, J. Tang, and J. Liu, "Temperature imaging of chips with high spatiotemporal resolution using diamond nitrogen-vacancy centers," Appl. Phys. Lett. **128**(9), 094001 (2026).

[8] J.W. Liu, M.Y. Liao, M. Imura, and Y. Koide, "Normally-off HfO2-gated diamond field effect transistors," Appl. Phys. Lett. **103**(9), 092905 (2013).

[9] T. Lamara, M. Belmahi, O. Elmazria, L. Le Brizoual, J. Bougdira, M. Rémy, and P. Alnot, "Freestanding CVD diamond elaborated by pulsed-microwave-plasma for ZnO/diamond SAW devices," Diam. Relat. Mater. **13**(4–8), 581–584 (2004).

[10] H. Nakahata, K. Higaki, S. Fujii, A. Hachigo, H. Kitabayashi, K. Tanabe, and S. Shikata, "SAW Devices on Diamond," in *1995 IEEE Ultrasonics Symposium. An International Symposium*, (Seatle, WA, USA, 1995), pp. 361–370.

[11] M.B. Assouar, O. Elmazria, P. Kirsch, P. Alnot, V. Mortet, and C. Tiusan, "High-frequency surface acoustic wave devices based on AlN/diamond layered structure realized using e-beam lithography," J. Appl. Phys. **101**(11), 114507 (2007).

[12] M.B. Schulz, B.J. Matsinger, and M.G. Holland, "Temperature dependence of surface acoustic wave velocity on α quartz," J. Appl. Phys. **41**(7), 2755–2765 (1970).

[13] N. Hara, M. Suzuki, S. Kakio, and Y. Yamamoto, "Analysis of longitudinal leaky surface acoustic waves on piezoelectric thin plates bonded to diamond substrate," Jpn. J. Appl. Phys. **62**, 1056 (2023).

[14] V. Mortet, O.A. Williams, and K. Haenen, "Diamond: A material for acoustic devices," Physica Status Solidi (A) Applications and Materials Science **205**(5), 1009–1020 (2008).

[15] M.W. Doherty, N.B. Manson, P. Delaney, F. Jelezko, J. Wrachtrup, and L.C.L. Hollenberg, "The nitrogen-vacancy colour centre in diamond," Phys. Rep. **528**(1), 1–45 (2013).

[16] F.R.L. Schoening, and L.A. Vermeulen, "X-Ray Measurements of Debye Temperature for Diamond at Low Temperatures," Solid State Commun. **7**(1), 15–18 (1969).

[17] J. Gugler, T. Astner, A. Angerer, J. Schmiedmayer, J. Majer, and P. Mohn, "Ab initio calculation of the spin lattice relaxation time T1 for nitrogen-vacancy centers in diamond," Phys. Rev. B **98**(21), 214442 (2018).

[18] A. Jarmola, V.M. Acosta, K. Jensen, S. Chemerisov, and D. Budker, "Temperature- and magnetic-field-dependent longitudinal spin relaxation in nitrogen-vacancy ensembles in diamond," Phys. Rev. Lett. **108**(19), 197601 (2012).

[19] A. Norambuena, E. Muñoz, H.T. Dinani, A. Jarmola, P. Maletinsky, D. Budker, and J.R. Maze, "Spin-lattice relaxation of individual solid-state spins," Phys. Rev. B **97**(9), 094304 (2018).

[20] B.A. McCullian, V. Sharma, H.Y. Chen, J.C. Crossman, E.J. Mueller, and G.D. Fuchs, "Coherent Acoustic Control of Defect Orbital States in the Strong-Driving Limit," PRX Quantum **5**(3), 030336 (2024).

[21] A. Albrecht, A. Retzker, F. Jelezko, and M.B. Plenio, "Coupling of nitrogen vacancy centres in nanodiamonds by means of phonons," New J. Phys. **15**, 083014 (2013).

[22] S.D. Bennett, N.Y. Yao, J. Otterbach, P. Zoller, P. Rabl, and M.D. Lukin, "Phonon-induced spin-spin interactions in diamond nanostructures: Application to spin squeezing," Phys. Rev. Lett. **110**(15), 156402 (2013).

[23] Z.E. Nataj, Y. Xu, D. Wright, J.O. Brown, J. Garg, X. Chen, F. Kargar, and A.A. Balandin, "Cryogenic characteristics of graphene composites—evolution from thermal conductors to thermal insulators," Nat. Commun. **14**(1), 3190 (2023).

[24] P.L. Kapitza, "Heat Transfer and Superfluidity of Helium II," Physical Review **60**, 354–355 (1941).

[25] E.T. Swartz, and R.O. Pohl, "Thermal boundary resistance," Rev. Mod. Phys. **61**, 605–668 (1969).

[26] A.A. Balandin, E.P. Pokatilov, and D.L. Nika, "Phonon Engineering in Hetero- and Nanostructures," Journal of Nanoelectronics and Optoelectronics **2**(2), 140–170 (2007).

[27] D.L. Nika, E.P. Pokatilov, and A.A. Balandin, "Phonon-engineered mobility enhancement in the acoustically mismatched silicon/diamond transistor channels," Appl. Phys. Lett. **93**(17), 173111 (2008).

[28] F. Kargar, and A.A. Balandin, "Advances in Brillouin–Mandelstam light-scattering spectroscopy," Nat. Photonics **15**(10), 720–731 (2021).

[29] P. Mutti, C.E. Bottani, G. Ghis, / Otti, M. Beghi, G.A.D. Briggs, and J.R. Sandercock, "Surface Brillouin Scattering-Extending Surface Wave Measurements to 20 GHz," in *Advances in Acoustic Microscopy*, (Springer, Boston, MA, 1995), pp. 249–300.

[30] J.R. Sandercock, "Trends in Brillouin Scattering: Studies of Opaque Materials, Supported Films, and Central Modes," in *Light Scattering in Solids III. Topics in Applied Physics*, (Springer, Berlin, Heidelberg, 1982), pp. 173–206.

[31] M.H. Grimsditch, and A.K. Ramdas, "Brillouin scattering in diamond," Phys. Rev. B **11**(8), 3139–3148 (1975).

[32] I. Motochi, S.R. Naidoo, B.A. Mathe, R. Erasmus, E. Aradi, T.E. Derry, and E.J. Olivier, "Surface Brillouin scattering on annealed ion-implanted CVD diamond," Diam. Relat. Mater. **56**, 6–12 (2015).

[33] I. Motochi, B.A. Mathe, S.R. Naidoo, and E. Aradi, "Anomalous behaviour of surface Brillouin scattering in thin strained CVD diamond," Diam. Relat. Mater. **109**, 108020 (2020).

[34] I. Motochi, B.A. Mathe, S.R. Naidoo, and T.E. Derry, "Surface Brillouin Scattering in Ion-implanted Chemical Vapor Deposited Diamond," in *Mater. Today Proc.*, (Elsevier Ltd, 2016), pp. S145–S152.

[35] Y.R. Xie, S.L. Ren, Y.F. Gao, X.L. Liu, P.H. Tan, and J. Zhang, "Measuring bulk and surface acoustic modes in diamond by angle-resolved Brillouin spectroscopy," Sci. China Phys. Mech. Astron. **64**(8), 287311 (2021).

[36] I. Motochi, B.A. Mathe, S.R. Naidoo, D. Wamwangi, and T.E. Derry, "Surface Brillouin scattering observation of higher order resonances in annealed, ion-implanted CVD diamond," Diam. Relat. Mater. **76**, 171–176 (2017).

[37] V. V Aleksandrov, T.S. Velichkina, J.B. Potapeva, and I.A. Yakovlev, "Mandelstamm-Brillouin studies of peculiarities of the phonon frequency distribution at cubic crystal (001) surfaces," Phys. Lett. A **171**(1–2), 103–106 (1992).

[38] P. Djemia, C. Dugautier, T. Chauveau, E. Dogheche, M.I. De Barros, and L. Vandenbulcke, "Mechanical properties of diamond films: A comparative study of polycrystalline and smooth fine-grained diamonds by Brillouin light scattering," J. Appl. Phys. **90**(8), 3771–3779 (2001).

[39] P. Djemia, C. Dugautier, T. Chauveau, E. Dogheche, M.I. De Barros, and L. Vandenbulcke, "Mechanical properties of diamond films: A comparative study of polycrystalline and smooth fine-grained diamonds by Brillouin light scattering," J. Appl. Phys. **90**(8), 3771–3779 (2001).

[40] A. Migliori, H. Ledbetter, R.G. Leisure, C. Pantea, and J.B. Betts, "Diamond's elastic stiffnesses from 322 K to 10 K," J. Appl. Phys. **104**(5), 053512 (2008).

[41] A. Nagakubo, M. Arita, H. Ogi, H. Sumiya, N. Nakamura, and M. Hirao, "Elastic constant C11 of 12C diamond between 10 and 613 K," Appl. Phys. Lett. **108**(22), 221902 (2016).

[42] H.J. McSkimin, and P. Andreatch, "Elastic moduli of diamond as a function of pressure and temperature," J. Appl. Phys. **43**(7), 2944–2948 (1972).

[43] E. Guzman, F. Kargar, A. Patel, S. Vishwakarma, D. Wright, R.B. Wilson, D.J. Smith, R.J. Nemanich, and A.A. Balandin, "Optical and acoustic phonons in turbostratic and cubic boron nitride thin films on diamond substrates," Diam. Relat. Mater. **140**, (2023).

[44] A.G. Every, "Pseudosurface wave structures in phonon imaging," Phys Rev B **33**(4), 2719–2732 (1986).

[45] N. Hara, M. Suzuki, S. Kakio, and Y. Yamamoto, "Analysis of longitudinal leaky surface acoustic waves on piezoelectric thin plates bonded to diamond substrate," Jpn. J. Appl. Phys. **62**, (2023).

[46] D. Wright, D.H. Mudiyanselage, E. Guzman, X. Fu, J. Teeter, B. Da, F. Kargar, H. Fu, and A.A. Balandin, "Acoustic and optical phonon frequencies and acoustic phonon velocities in Si-doped AlN thin films," Appl. Phys. Lett. **125**(14), 142202 (2024).

[47] D. Wright, E. Guzman, M.S.H. Bijoy, R.B. Wilson, D.H. Mudiyanselage, H. Fu, F. Kargar, and A.A. Balandin, "Acoustic phonon characteristics of (001) and (201) β-Ga2O3 single crystals investigated with Brillouin-Mandelstam light scattering spectroscopy," Appl. Phys. Lett. **127**(5), 052201 (2025).

[48] H.R. Phillip, and E.A. Taft, "Kramers-Kronig Analysis of Reflectance Data for Diamond," Physical Review **136**(5A), 1445–1448 (1964).

[49] M.N. Polyanskiy, "Refractiveindex.info database of optical constants," Sci. Data **11**(94), 1–19 (2024).

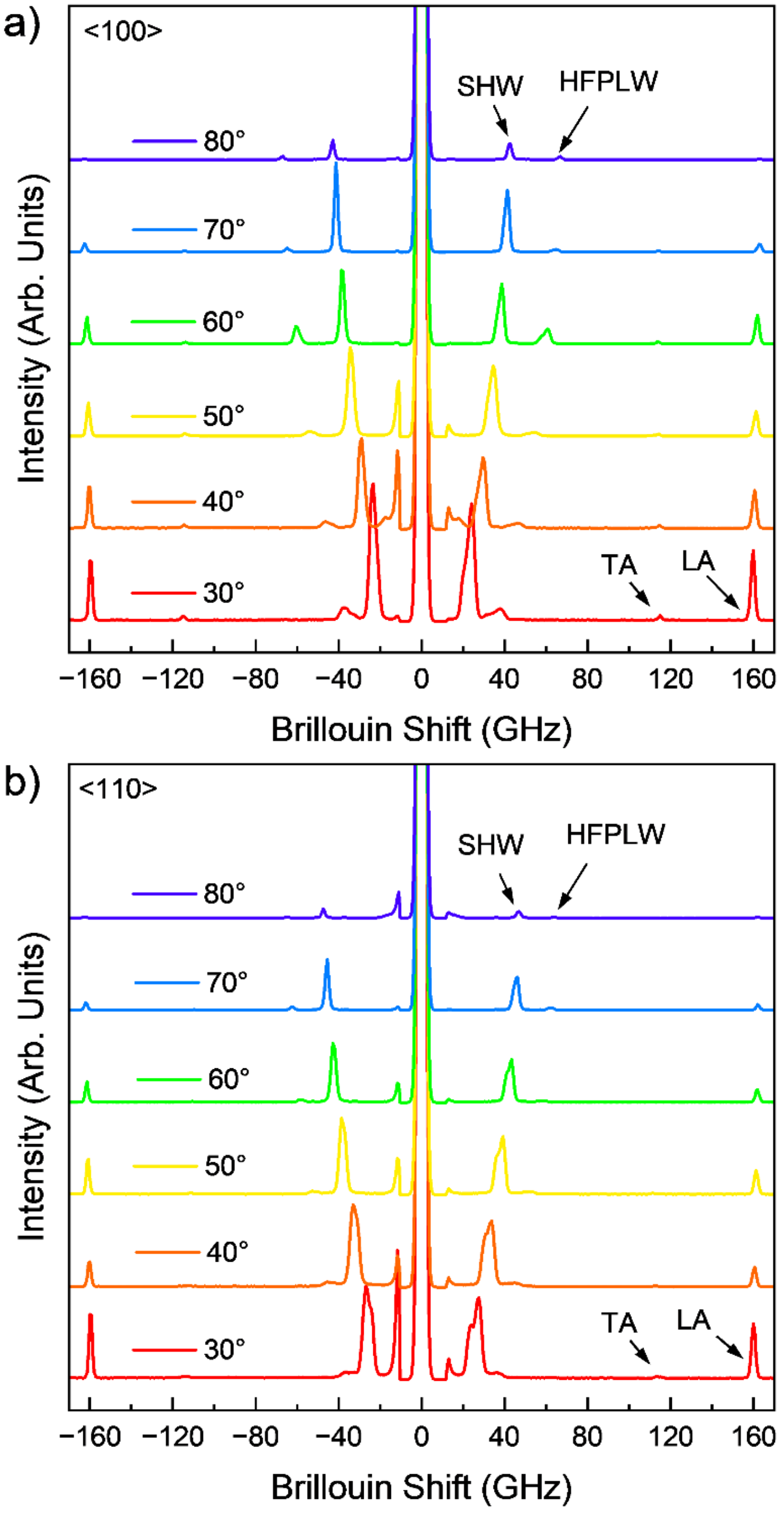
a)
<100>
SHW
HFPLW
80°
70°
60°
50°
40°
30°
TA
LA
Intensity (Arb. Units)
−160
−120
−80
−40
0
40
80
120
160
Brillouin Shift (GHz)
b)
<110>
SHW
HFPLW
80°
70°
60°
50°
40°
30°
TA
LA
Intensity (Arb. Units)
−160
−120
−80
−40
0
40
80
120
160
Brillouin Shift (GHz)

Figure 1

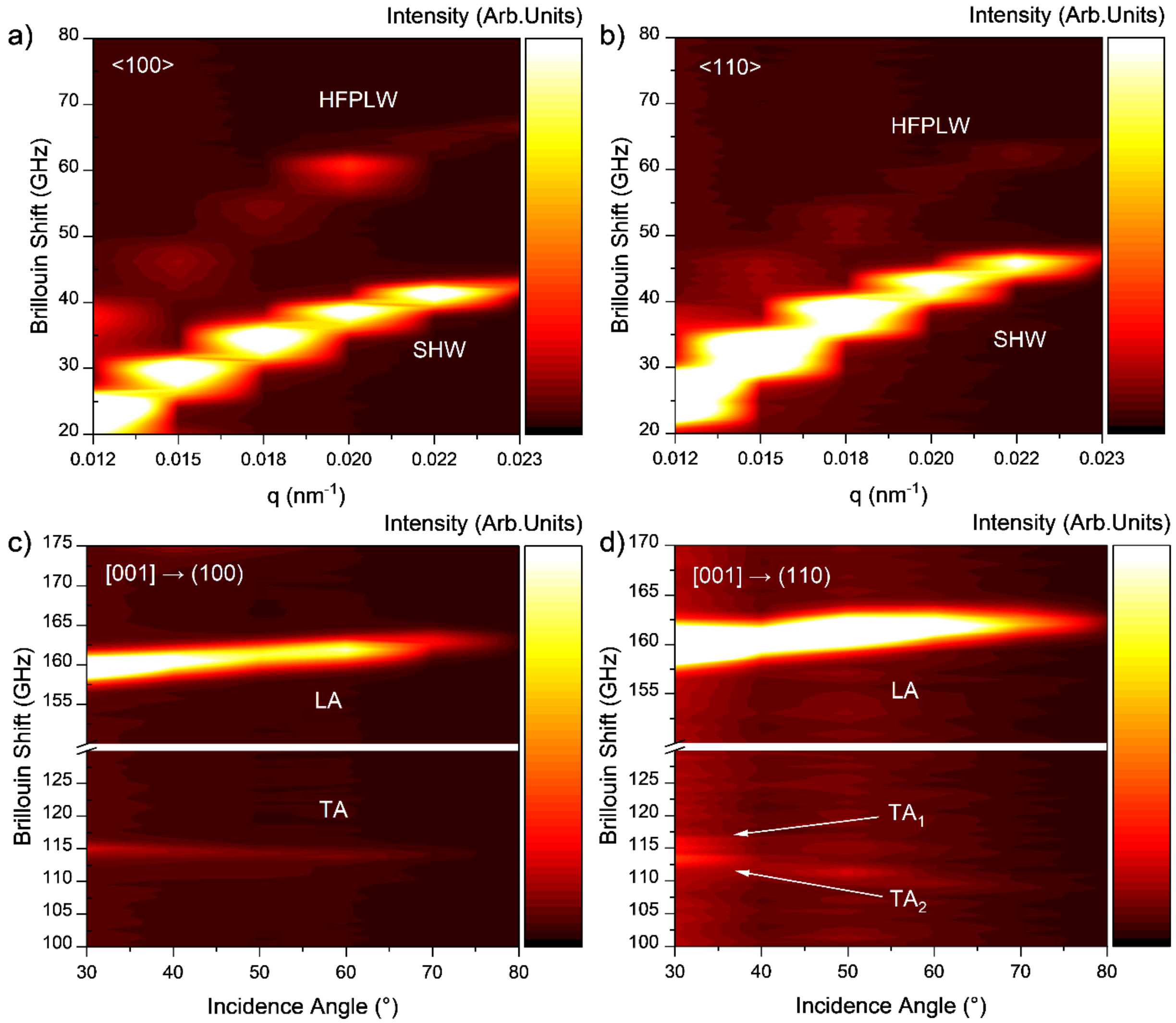

a)
Intensity (Arb.Units)
<100>
HFPLW
SHW
Brillouin Shift (GHz)
q (nm$^{-1}$)
b)
Intensity (Arb.Units)
<110>
HFPLW
SHW
Brillouin Shift (GHz)
q (nm$^{-1}$)
c)
Intensity (Arb.Units)
[001] → (100)
LA
TA
Brillouin Shift (GHz)
Incidence Angle (°)
d)
Intensity (Arb.Units)
[001] → (110)
LA
TA$_1$
TA$_2$
Brillouin Shift (GHz)
Incidence Angle (°)

Figure 2

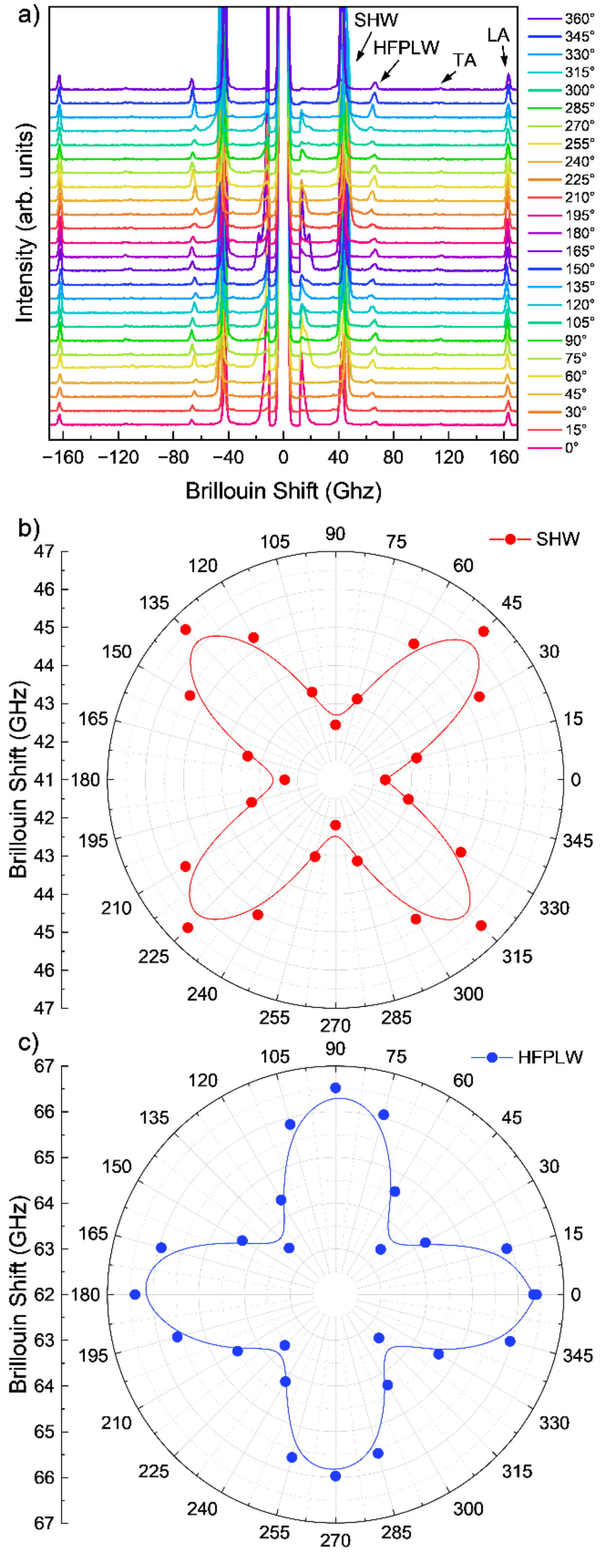
a)
SHW
HFPLW
TA
LA
Intensity (arb. units)
Brillouin Shift (Ghz)
b)
SHW
Brillouin Shift (GHz)
c)
HFPLW
Brillouin Shift (GHz)

Figure 3

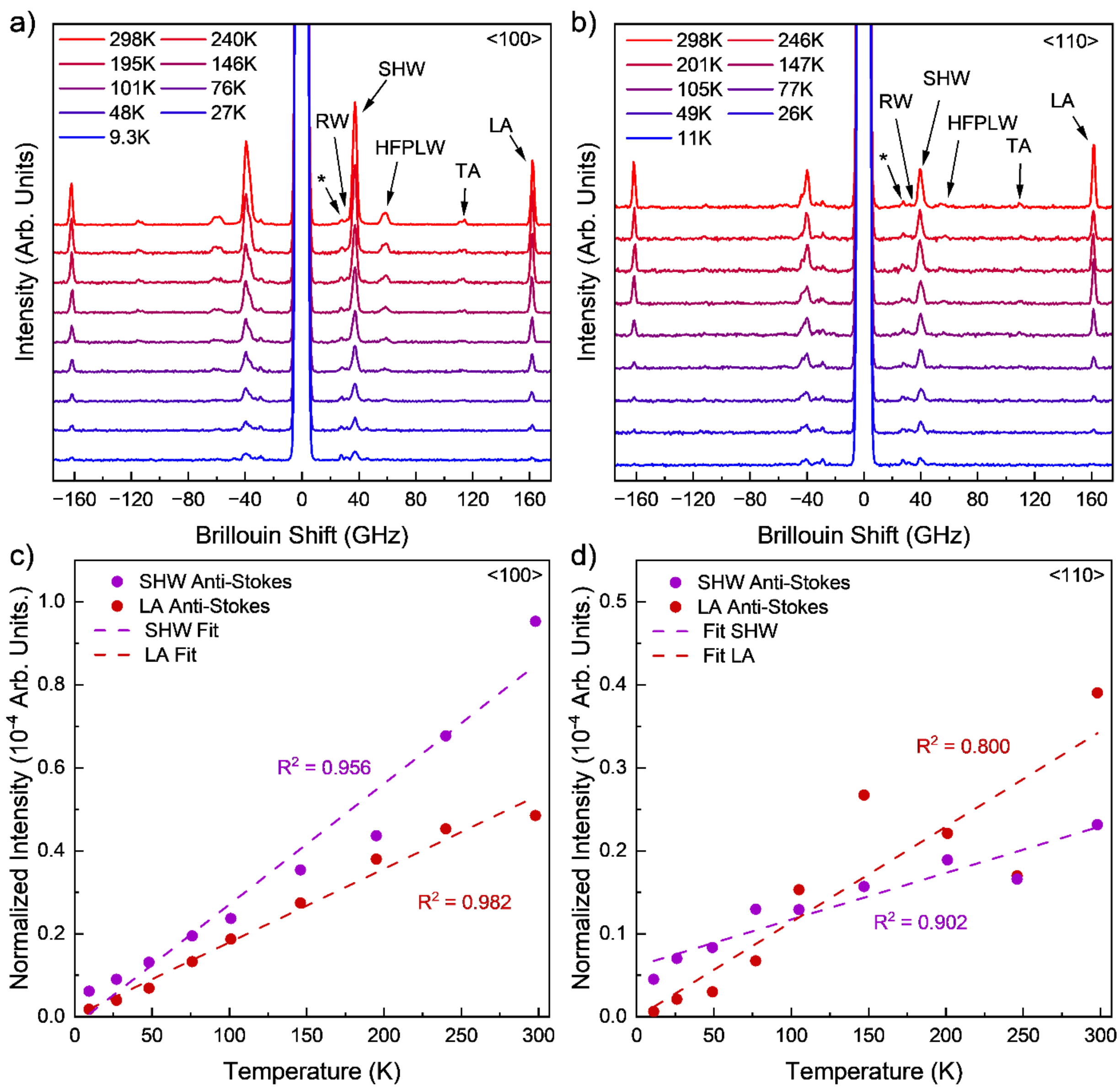

a)
298K
240K
195K
146K
101K
76K
48K
27K
9.3K
<100>
SHW
RW
*
HFPLW
TA
LA
Intensity (Arb. Units)
−160 −120 −80 −40 0 40 80 120 160
Brillouin Shift (GHz)
b)
298K
246K
201K
147K
105K
77K
49K
26K
11K
<110>
SHW
RW
*
HFPLW
TA
LA
Intensity (Arb. Units)
−160 −120 −80 −40 0 40 80 120 160
Brillouin Shift (GHz)
c)
SHW Anti-Stokes
LA Anti-Stokes
SHW Fit
LA Fit
<100>
$R^2$ = 0.956
$R^2$ = 0.982
Normalized Intensity ($10^{-4}$ Arb. Units.)
0.0 0.2 0.4 0.6 0.8 1.0
0 50 100 150 200 250 300
Temperature (K)
d)
SHW Anti-Stokes
LA Anti-Stokes
Fit SHW
Fit LA
<110>
$R^2$ = 0.800
$R^2$ = 0.902
Normalized Intensity ($10^{-4}$ Arb. Units.)
0.0 0.1 0.2 0.3 0.4 0.5
0 50 100 150 200 250 300
Temperature (K)

Figure 4

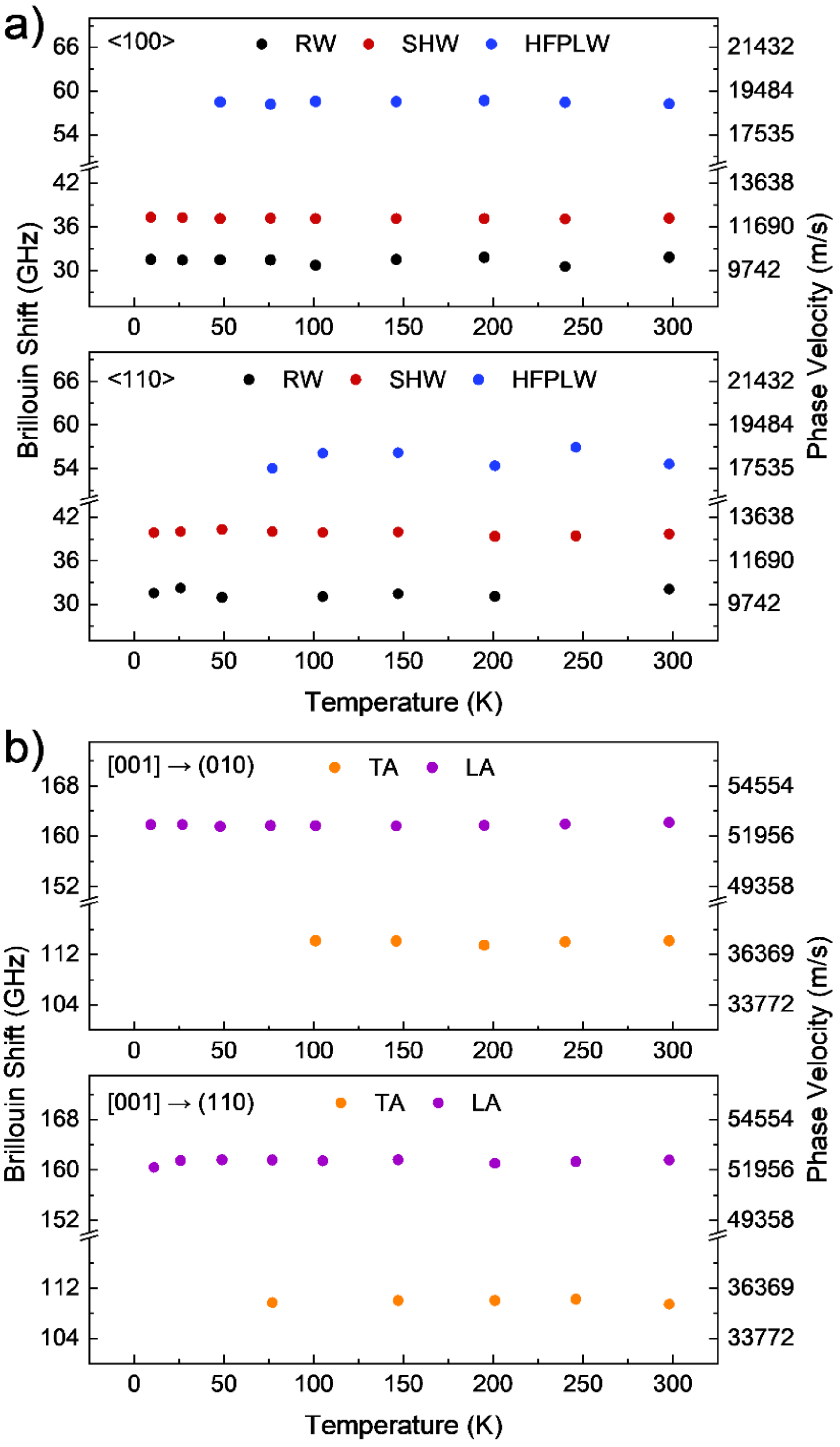

a)
<100>
RW
SHW
HFPLW
<110>
Brillouin Shift (GHz)
Phase Velocity (m/s)
Temperature (K)
b)
[001] → (010)
TA
LA
[001] → (110)

Figure 5